\documentclass[12pt,leqno,twoside]{article}
\usepackage{pdfsync}
\usepackage{mathrsfs,amssymb,amsfonts,amsthm,amsmath,natbib,amsbsy, bbm, bm}
\usepackage{dsfont,verbatim,fancyhdr}
\usepackage{graphicx,float} 
\usepackage[hang]{caption} 
\usepackage[usenames,dvipsnames]{xcolor}
\usepackage[bookmarks,bookmarksnumbered,colorlinks=true,pdfstartview=FitV, linkcolor=Red,citecolor=blue!90!green,urlcolor=blue!90!green]{hyperref}
\usepackage{tkz-graph}
\usepackage{tikz}
\usepgflibrary{arrows}
\usepackage{cancel}
\usepackage{subcaption}
\allowdisplaybreaks

\bibpunct{\textcolor{blue!90!green}{(}}{\textcolor{blue!90!green}{)}}{\textcolor{blue!90!green}{;}}{a}{\textcolor{blue!90!green}{,}}{\textcolor{blue!90!green}{;}}

\newcommand{\bfk}[0]{\mathbf{k}}
\newcommand{\bfn}[0]{\mathbf{n}}

\newcommand{\bfm}[0]{\mathbf{m}}
\newcommand{\MM}[0]{\mathbf{M}}
\newcommand{\bfM}[0]{\mathcal{M}}

\newcommand{\bbY}[0]{\mathbb{Y}}
\newcommand{\bfY}[0]{\mathbf{Y}}
\newcommand{\bfalpha}[0]{\boldsymbol{\alpha}}

\newcommand{\bfkpast}[0]{\mathbf{k}_{i-1}}
\newcommand{\bfkfuture}[0]{\mathbf{k}_{i+1}}
\newcommand{\bfnpast}[0]{\mathbf{n}_{i-1}}
\newcommand{\bfnfuture}[0]{\mathbf{n}_{i+1}}
\newcommand{\bfmpast}[0]{\mathbf{m}_{i-1}}
\newcommand{\bfmfuture}[0]{\mathbf{m}_{i+1}}
\newcommand{\Z}[0]{\mathbb{Z}}

\newcommand{\N}[0]{\mathbb{N}}

\newcommand{\Law}[0]{\mathcal{L}}

\newcommand{\iid}{\overset{\text{iid}}{\sim}}

\newcommand*\diff{\mathop{}\!\mathrm{d}}

\newtheorem{theorem}{Theorem}[section]
\newtheorem{proposition}[theorem]{Proposition}

\newtheorem{definition1}[theorem]{Definition} 
 
\newtheorem{remark1}[theorem]{Remark} 
\long\def\symbolfootnote[#1]#2{\begingroup\def\thefootnote{\hspace*{-1mm}\fnsymbol{footnote}}\footnote[#1]{#2}\endgroup}

\title{\bf %\vspace{-2.5cm} 
An R package for nonparametric inference on dynamic populations with infinitely many types
}

\author{
\textsc{\normalsize Filippo Ascolani}, \emph{\normalsize Duke University}\\[2mm]
\textsc{\normalsize Stefano Damato}, \emph{\normalsize USI-SUPSI, Switzerland}\\[2mm]
\textsc{\normalsize Matteo Ruggiero},
\emph{\normalsize University of Torino and Collegio Carlo Alberto}
}

\date{ \today}

\begin{document}

\maketitle
\thispagestyle{empty}

\vspace{-5mm}
\begin{center}
\begin{minipage}{.75\textwidth}
\footnotesize\noindent
Fleming--Viot diffusions are widely used stochastic models for population dynamics which extend the celebrated Wright--Fisher diffusions. They describe the temporal evolution of the relative frequencies of the allelic types in an ideally infinite panmictic population, whose individuals undergo random genetic drift and at birth can mutate to a new allelic type drawn from a possibly infinite potential pool, independently of their parent. Recently, Bayesian nonparametric inference has been considered for this model when a finite sample of individuals is drawn from the population at several discrete time points. Previous works have fully described the relevant estimators for this problem, but current software is available only for the Wright--Fisher finite-dimensional case. Here we provide software for the general case, overcoming some non trivial computational challenges posed by this setting. \\
The R package FVDDPpkg efficiently approximates the filtering and smoothing distribution for Fleming--Viot diffusions, given finite samples of individuals collected at different times. A suitable Monte Carlo approximation is also introduced in order to reduce the computational cost. \\
Software available at \href{https://github.com/StefanoDamato/FVDDPpkg}{https://github.com/StefanoDamato/FVDDPpkg}.
\\[-2mm]


%\textbf{MSC} Primary:
%62M20, % prediction/filtering
%%60G35, % Signal detection and filtering
%%93E11, % Filtering
%Secondary:
%%94A12, % Signal theory (characterisation, reconstruction, filtering, etc.)
%60J60. % diffusion processes
%92D25, % pop dynamics
%60G57. %random measure
\end{minipage}
\end{center}

\vfill
\newpage

%%%%%%%%%%%%%%%%%%%%%%%%%
%%%%%%%%%%%%%%%%%%%%%%%%%%
\section{Introduction}

Fleming--Viot (FV) diffusions are among the most widely used models in population genetics and stochastic population dynamics. See \cite{EthierKurtz1993} for an extensive review. They take values in the space of purely atomic probability measures defined on an arbitrary complete and separable metric space $\bbY$, which in our application denotes the set of all possible \emph{allelic types}. The process, denoted $\{X_t,t\ge0\}$, describes the temporal evolution of the relative frequencies of types in an ideally infinite population, when the individuals that compose the population undergo random genetic drift, due to panmictic random mating, and at birth can mutate to a yet unseen type, drawn from a possibly infinite potential pool. Fleming--Viot diffusion can also accommodate mathematical descriptions of other evolutionary forces like natural selection or recombination, but in this paper we do not consider these cases. 
The time-homogeneous transition function, found by \cite{EthierGriffiths1993}, can be written
\begin{equation}\label{eq:FVtransition}
  P_t(x, \diff x') = \sum_{m=0}^\infty d^{\theta}_m(t) \int_{\bbY^m} \Pi_{\alpha + \sum_{k=1}^m \delta_{y_k}} (\diff x') x(\diff y_1)\cdots x(\diff y_m).
\end{equation}
See eq.~(2) in the Supplementary Material for the diffusion infinitesimal generator.
Here $\Pi_\alpha$ is the reversible measure of the FV diffusion and denotes the law of a Dirichlet process \citep{Ferguson1973} with baseline measure $\alpha = \theta P_0$, with $\theta>0$ being the overall mutation rate and $P_{0}$ a probability measure on $\bbY$ representing the potential pool of allelic types from which mutants are drawn. Furthermore, $d^{\theta}_m(t)$ is the transition probability of a pure-death process on $\mathbb{Z}_+$ which starts at infinity almost surely and jumps from $m$ to $m-1$ at rate $\lambda_m = m(\theta+m-1)/2$.  These were computed in \cite{GRIFFITHS198037} and \cite{Tavare1984} and are in fact the transition probabilities of the block-counting process in Kingman's coalescent with mutation, given by
\begin{equation}\label{BC process trans probabilities infty}
d_{m}^\theta(t) =
\left\{
\begin{array}{ll}
 \sum_{k\ge m}e^{-\lambda_{k}t}
(-1)^{k-m}\binom{k}{m}
\frac{(\theta+2k-1)(\theta+m)_{(k-1)}}{k!},\quad  &  m\ge1, \\
1- \sum_{k\ge 1}e^{-\lambda_{k}t}
(-1)^{k-1}
\frac{(\theta+2k-1)(\theta)_{(k-1)}}{k!},\quad  &  m=0, 
\end{array}
\right.
\end{equation}
where $(a)_{(b)} = a(a+1)\dots(a+b-1)$ is the ascending factorial or Pochhammer symbol. A different representation of \eqref{eq:FVtransition} highlights the role of the death process as latent variable which determines the correlation between $X_{s}$ and $X_{s+t}$ (here $s$ can be set equal to 0 as the process is time-homogeneous), and reads
\begin{equation}\label{eq:FVtransition_alternative}
\begin{aligned}
X_{t}\mid & (D_{t}=m,Y_{0,1},\ldots,Y_{0,m})\,\sim \, \Pi_{\alpha+ \sum_{i = 1}^m\delta_{Y_{0,i}}}
\quad \quad 
Y_{0,i}\mid X_{0}\overset{\text{iid}}{\sim }X_{0}.
\end{aligned}
\end{equation}
Here $D_t$ is the above death process with transition probabilities $d^{\theta}_m(t)$. The realisation of $D_{t}$ determines the number $m\in \Z_{+}$ of individuals whose type is drawn from the initial state $X_{0}$ and used in the distribution of the arrival state $X_t$. The intuition is that for small $t$ a large $m$  is more likely, so that $X_{0}$ and $X_{t}$ will tend to be similar, whereas at $t\rightarrow \infty$, $m$ will be small or zero with high probability, and $X_{0}$ and $X_{t}$ will be essentially uncorrelated. Cf.~also \cite{walker2007fleming}. 

%Indeed it can be shown that the FV processes arise as an appropriate limit, as the number of individuals of the population grows, of the frequencies associated to the Wright-Fisher model in population genetics (see Chapter $4$ in \cite{ethier1986markov} or \cite{etheridge2011some}).

A projection of the above FV diffusion onto a measurable partition $A_{1},\ldots,A_{K}$ of the type space $\bbY$ yields a \emph{neutral} Wright--Fisher (WF) diffusion with $K$ types, cf.~\cite{Etheridge2000} and \cite{Feng2010}. Recently, \cite{JenkinsSpano2017} and \citet{sant_ewf_2023} addressed the problem of generating exact draws from the  transition functions of this and other WF diffusions, in particular addressing the intractability of \eqref{BC process trans probabilities infty}, which also appears in the associated finite-dimensional special case of \eqref{eq:FVtransition}.

Given the increasing availability of population-level allelic data, it is often of interest to study the temporal evolution of the associated frequencies, which can be modelled through a FV or a WF diffusion. See, e.g., \cite{tataru2017statistical},\citet{gory2018bayesian},\ and \citet{paris2019inference} for recent works in this direction.
From a statistical perspective, the problem can be formulated as a hidden Markov model, a popular inferential framework for temporally evolving quantities \citep{CappeMoulinesRyden2005}. In such setting, the FV process which models the evolving allelic frequencies is taken as the unobserved target of inference, and  finitely-many biological samples from the population are assumed to be collected at discrete times. Denote by $Y_{t_k, i}$ the type of the $i$th individual observed at time $t_{k}$, {for $i=1,\ldots,n_{k}$} and times $0 = t_0 < t_1 < \dots < t_{p} = T$. Given a dataset $\bfY_{0:T}:=(\bfY_{0},\ldots,\bfY_{T})$, where $\bfY_{k}:=(Y_{t_{k},1},\ldots,Y_{t_{k},n_{k}})$ is a row vector, the task is then to make inference on the process states, hence on the configuration of frequencies in the underlying population. Mathematically this amounts to fully describe the law of $X_t \mid \bfY_{0:T}$, which is typically called  \emph{filtering distribution} when $t= T$, or \emph{smoothing distribution} when $0 \leq t < T$. 
See Section \ref{sec:model} for a more detailed formalization of the setting. These inferential problems were fully solved and implemented by 
\cite{PapaspiliopoulosRuggiero2014} and \citet{kon2021exact,kon2024}
for the finite-dimensional WF special case. 
In the above infinite-dimensional framework, \cite{PapaspiliopoulosRuggieroSpano2016}, \citet{AscolaniLijoiRuggiero2021} and \citet{AscolaniLijoiRuggiero2023} provided recursions that in principle allow to fully evaluate such distributions, showing that they are qualitatively similar to the formulation \eqref{eq:FVtransition_alternative}, with the important difference that the latent variable $D_{t}$ is substituted by another variable with \emph{finite state space}, yielding in fact finite mixtures of laws of Dirichlet processes structurally similar to $\sum_{i\in I}v_{i}\Pi_{\alpha_{i}}$, where $I$ has finite cardinality and $\sum_{i\in I}v_{i}=1$ (see Theorem $1$ in the Supplementary Material). These results surprisingly overcome the intractability of the death process $D_{t}$ in \eqref{eq:FVtransition_alternative}, and in particular the need to deal with the infinite series expansion for its transition probabilities, making the evaluation of these distributions possible with a finite computational effort, without the need for stochastic truncation or approximation strategies. 
However, sampling in practice from these distributions leads to non trivial computational challenges that are not present in the finite-dimensional subcase, mainly related to the growth in the number of observed types  as more data are collected. This aspect reflects the growth of the cardinality of $I$ in the above expression, making exact computations quickly infeasible. Another aspect which is peculiar of this setting is how the weights $v_{i}$ in some of the distributions of interest depend on the support of the offspring distribution $P_0$, which from a modelling perspective can be taken as atomic or nonatomic.

Here we present a simple and efficient Monte Carlo procedure to approximate the filtering and smoothing distributions of a FV process under the assumed data collection framework. These are implemented in the publicly available R package \emph{FVDDPpkg}.
The acronym for the package relates to another line of research on \emph{dependent Dirichlet processes} (DDPs), which in Bayesian nonparametric statistics are considered as collections of correlated random probability measures with Dirichlet process marginals. See \cite{QuintanaJaraMacEachern2022} for a recent review.

\section{Model}\label{sec:model}

We assume a hidden Markov model formulation, whereby the unobserved process (sometimes called \emph{signal}) $X_{t}$ is a FV diffusion with dynamics as in \eqref{eq:FVtransition}, denoted $X \sim \text{FV}(\alpha)$, and the observations, which provide the allelic  types of the individuals, are conditionally \emph{iid} given the current FV state, namely $Y_{t_{k},i} | X_{t_{k}} = x \overset{\text{iid}}{\sim} x$. That is, when the current configuration of the type frequencies is the atomic probability measure $x$ on $\bbY$, so that this admits representation $x=\sum_{i=1}^{\infty}v_{i}\delta_{z_{i}}$, with $\sum_{i\ge1}v_{i}=1$ and $z_{i}\in \bbY$, then $Y_{t_{k},i}=z_{i}$ with probability $v_{i}$. This essentially amounts to sampling without replacement when the population from which we draw is infinite.

Let now $X_0 \sim \Pi_\alpha$ be the initial state of the process. Recall that $\alpha = \theta P_0$, with $\theta>0$ a fixed constant and $P_{0}$ a probability measure on $\bbY$. Denote as above by $\bfY_{0:T}:=(\bfY_{0},\ldots,\bfY_{T})$ the overall dataset, where $\bfY_{k}:=(Y_{t_{k},1},\ldots,Y_{t_{k},n_{k}})$ for times $0 = t_0 < t_1 < \dots < t_{p} = T$, and let $K$ be the number of unique values (allelic types) in $\bfY_{0:T}$. We are interested in the \emph{filtering} and \emph{smoothing distribution} for $X_t$, which can be written \citep{PapaspiliopoulosRuggieroSpano2016} and \citet{AscolaniLijoiRuggiero2023} as
\begin{equation}\label{smoothing_distribution}
X_t | (\bfY_{0:T},\MM_{t}=\bfm) \sim  \Pi_{\alpha+\sum_{k = 1}^Km_k\delta_{y_k^*}},\quad \quad 
p(\MM_{t}=\bfm)=w_\bfm(t).
\end{equation}
Here $\MM_{t}$ is a certain death process on a finite state space $\bfM\subset \Z_{+}^K$, where $\bfm = (m_1, \dots, m_K)\in \bfM$ describes the multiplicities of $(y_1^*, \dots, y_K^*)$, the unique values (types) in $\bfY_{0:T}$. These elements input information in the mixture components through the empirical measure $\sum_{k = 1}^Km_k\delta_{y_k^*}$.  Notice that \eqref{smoothing_distribution} is structurally similar to representation \eqref{eq:FVtransition_alternative}, with the key difference that the death process $\MM_{t}$ has a \emph{finite} state space. 
Further details on this representation, including $\bfM$ and $w_\bfm(t)$, can be found in the Supplementary Material.

These distributions allow one to make inference on the evolution of the frequencies in the population, given the data, at any time of interest. Moreover, they can be used to evaluate the probability of observing again the already recorded allelic types in new individuals drawn from the population; see Proposition $2$ in \cite{AscolaniLijoiRuggiero2021} and Section $1$ in the Supplementary Material.

\section{Methods}\label{methods}

The task of evaluating the right hand side of \eqref{smoothing_distribution} poses significant challenges. The probability weights $w_\bfm(t)$ depend on the transition probabilities of a certain death process (informally, a projection of $D_{t}$ in \eqref{eq:FVtransition}), whose computation can be numerically unstable and requires some care (\cite{griffiths1984asymptotic}). Even assuming regularly spaced data collection times at intervals of length $\Delta$, the cardinality of $\bfM$ grows polynomially with the number of observed datapoints (cf.~Section $3.2$ in \cite{AscolaniLijoiRuggiero2023}), thus storing all the $w_\bfm(\Delta)$ becomes demanding even with moderately large datasets. Finally, when $t < T$ as in \eqref{smoothing_distribution}, the set of nodes $\bfm$ with strictly positive probability depends on the nature of baseline distribution $P_0$ (Proposition $3.6$ in \cite{AscolaniLijoiRuggiero2021}). Specifically, if $P_{0}$ is nonatomic, $\bfM$ depends on the unique values shared across different data collection times (cf.~point C of Proposition $3.6$ in \cite{AscolaniLijoiRuggiero2021}).

An accurate approximation can nevertheless be obtained by exploiting different factors. First, the weights can be computed  recursively. If $t \in (t_{j}, t_{j+1})$, \eqref{smoothing_distribution} can be computed starting by the distributions of the signal given past and future observations used separately, i.e.
\begin{equation}\label{past_future}
\begin{aligned}
\mathcal{L}&\,\left(X_{t_j}  | \bfY_{0:j}\right) = \sum_{\bfm_1 \in \bfM_1}u_{\bfm_1}\Pi_{\alpha+\sum_{k = 1}^Km_{1,k}\delta_{y_k^*}},\\ 
\mathcal{L}&\,\left(X_{t_{j+1}}  | \bfY_{j+1: T}\right) = \sum_{\bfm_2 \in \bfM_2}v_{\bfm_2}\Pi_{\alpha+\sum_{k = 1}^Km_{2,k}\delta_{y_k^*}}, 
\end{aligned}
\end{equation}
where $\bfM_1 \subset \N^K$ and $\bfM_2 \subset \N^K$ are suitable sets
(see Sections $2$ and $3$ of the Supplementary material). 
Below we focus on the (arguably harder) case of approximating the smoothing distribution, while the procedure for filtering is discussed in Section $4$ of the Supplementary material. Each weight $w_\bfm$ in \eqref{smoothing_distribution} for the smoothing distribution turns out to be
\begin{equation}\label{weights}
w_\bfm = 
\sum_{\bfm_1 \in \bfM_1}
\sum_{ \bfm_2 \in \bfM_2}
u_{\bfm_1}v_{\bfm_2}\sum_{\bfk_1, \bfk_2 \, : \, \bfk_1+ \bfn +  \bfk_2 = \bfm}\tilde{w}^{\bfm_1, \bfm_2}_{\bfk_1, \bfn, \bfk_2},
\end{equation}
with $u_{\bfm_1},v_{\bfm_2}$ taken from  \eqref{past_future} and where
\[
\tilde{w}^{\bfm_1, \bfm_2}_{\bfk_1, \bfn, \bfk_2} \propto p_{\bfm_1, \bfk_1}(t-t_i)p_{\bfm_2, \bfk_2}(t_{i+1}-t)q(\bfk_1, \bfn, \bfk_2).
\]
Here in turn the vector indices $\bfn$ are the multiplicities of the allelic types observed at the time $t$ of interest (one can simply set $\bfn$ to be a vector of zeros if at time $t$ no data are available), $p_{\bfn, \bfk}(t)$ are the transition probabilities of a death process on $\Z_{+}^K$ which decreases from $\bfm \in \Z_{+}^K$ to $\bfm-\textbf{e}_j$ at rate $\lambda_{\bfm, j} := m_j(|\bfm|+\theta-1)$, where $\textbf{e}_j$ denotes the canonical  vector in the direction $j$, and $q$ is a suitable function which depends on $P_0$. Weights in \eqref{weights} represent the probability that, starting two independent death processes at the points of $\bfM_1$ and $\bfM_2$, the arrival points $\bfk_1$ and $\bfk_2$ sum up to $\bfm - \bfn$. This is then reweighted through the function $q$, which formalizes how the information contained in the data collected prior to and after $t$ is combined. 
Further details are given in Section 3 of the Supplementary Material (cf.~Proposition~5).

Computing the transition probabilities $p_{\bfn, \bfk}(t)$ is the hardest task. Indeed, their exact value depends on a sum with alternating signs (see formula (B40) in \cite{kon2021exact}) and they assign most of the probability mass to few nodes, hence a large majority of elements in $\bfM$ retains a low cumulative probability (cf.~Figure $4$ in \cite{AscolaniLijoiRuggiero2021}). \cite{kon2021exact} (cf.~Section 4.2) exploit this principle for Wright--Fisher diffusions through a pruning strategy. Here we instead consider a Monte Carlo approximation of  \eqref{weights}, leading to the following key part of  the algorithm for approximating the terms $p_{\bfn, \bfk}(t)$. Given \eqref{past_future}:
\begin{list}{
$\bullet$
}{\itemsep=1mm\topsep=2mm\itemindent=-3mm\labelsep=2mm\labelwidth=0mm\leftmargin=9mm\listparindent=0mm\parsep=0mm\parskip=0mm\partopsep=0mm\rightmargin=0mm\usecounter{enumi}}
\setcounter{enumi}{0}
\item draw $\bfm_1$ and $\bfm_2$  with probability $u_{\bfm_1}$ and $v_{\bfm_2}$ respectively;
\item simulate two independent death processes with rates $\lambda_{\bfm_1}$ and $\lambda_{\bfm_2}$ over the time intervals $t-{t_{j}}$ and $t_{j+1}-t$ respectively, where $\lambda_{\bfm}:=(\lambda_{\bfm, 1},\ldots,\lambda_{\bfm, K})$ and $\lambda_{\bfm, j}$ is as above;
\item return the two arrival nodes $\bfk_1$ and $\bfk_2$.
\end{list}

The death process simulation can be performed by means of the Gillespie algorithm \citep{gillespie2007stochastic}, as described in the Supplementary Material.
Once the above procedure is repeated for the desired number of particles, for a fixed $\bfm$ the unnormalized version of $w_\bfm$ is approximated by the empirical average of $q\left(\bfk_1, \bfn, \bfk_2 \right)$, for all the sampled $(\bfk_1, \bfk_2)$ such that $\bfm = \bfk_1+ \bfn + \bfk_2$. Normalizing the weights is straightforward as their cardinality is finite, and after normalization we can in addition apply a suitable \emph{pruning}, that is all the weights below a certain threshold (which we chose no larger than $10^{-9}$) are eliminated and a new normalization is performed. Further details can be found in Section 4 of the Supplementary Material.

\begin{figure}[t!]
\centering
\includegraphics[width=0.8\textwidth]{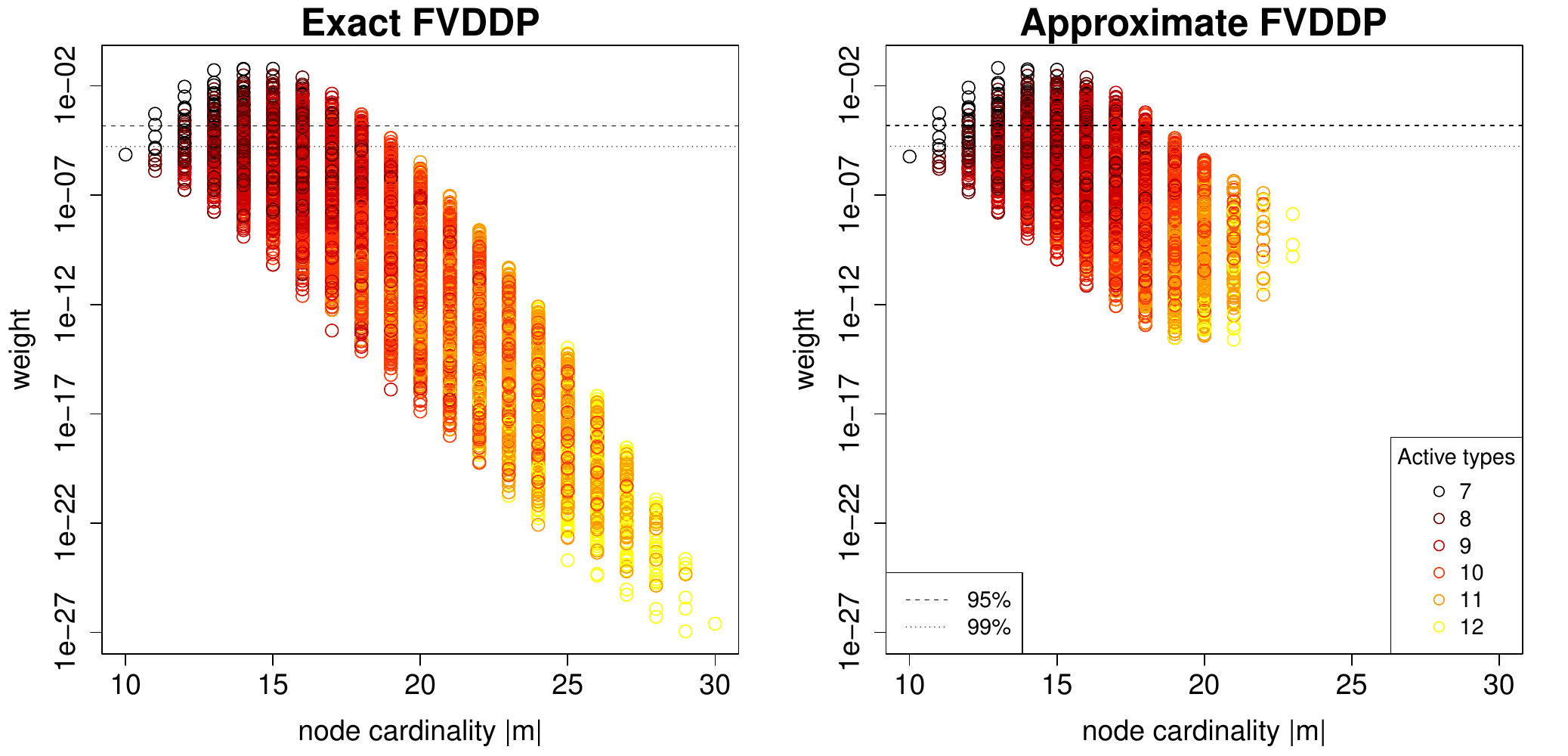}

\begin{quote}

    \caption{
    \footnotesize
    Comparison between probability mass assigned to mixture components in the exact (left) and approximate (right) smoothing distribution. The simulation is based on synthetic data from $3$ different times, each with $10$ observations, and $10^6$ particles for the death process simulations.
The cardinality $|\bfm |= \sum_{j=1}^Km_j$ associated to each node is depicted on the $x$-axis, with the corresponding probability weight (in log scale) on the y-axis. Observations are color-coded based on the amount of active types ($\mathrm{card}\{ j: m_j \neq 0 \}$), darker colors indicating lower values. Above the dashed lines lie the $95\%$ and $99\%$ of the total mass, respectively.  The figure shows that the exact distribution can be efficiently approximated through simulation of the death processes by mixture  with a significant lower number of components having non negligible probability mass.
    }\label{fig1}
\end{quote}
\end{figure}

The above strategy typically leads to mixtures with a relatively small number of components which collectively provide a good approximation to the true distribution. 
Figure \ref{fig1} shows that the algorithm is able to reproduce efficiently large mixtures. The exact smoothing distribution in the left panel is composed of more than $30.000$ components, many of which have an almost null weight: indeed $90\%$, $95\%$, and $99\%$ of the mass is given by respectively $237$, $432$, and $1170$ nodes with the largest weight. The approximate distribution whose nodes are represented in the right panel captures the same information using $12.500$ components, with an average absolute error of order $5 \cdot 10^{-6}$. In the figure, the $x$-axis depicts the cardinality of the vectors $\bfm$, showing that the vast majority of the probability mass is retained by nodes with few elements: this is coherent with the fact that the death process $\MM_t$ decreases very fast (cf.~Figure $4$ in \cite{AscolaniLijoiRuggiero2021}). See Section 6 of the Supplementary Material for further details on the simulation setting and the associated computational cost: in particular, Figure $3$ in the Supplementary Material depicts the growth of the (time and memory) cost as a function of the Monte Carlo samples. Clearly, one could choose a higher threshold value for pruning the mixture components, leading to more parsimonious mixtures and higher computational efficiency, at the cost of a larger approximation error. The feather-like shape in the figure shows that nodes corresponding to higher cardinalities have typically smaller weights. This behaviour is mainly a product of the fact that the death process propagates the probability mass towards the empty node (the origin of the graph in Figure 1 in the Supplementary Material). Hence the interplay between updates with new data, which moves the mass upwards (cf.~Figure 1 in the Supplementary Material), with the temporal propagation which operates for both distributions in \eqref{past_future}, which become part of the smoothing distribution, ends up keeping most of the probability mass in nodes that are relatively near the origin.

\newpage

\newpage
\section{Discussion}

The $R$ package \emph{FVDDPpkg} allows one to compute the filtering and smoothing distributions for the above FV-driven hidden Markov model, with an arbitrary set of observation times. It also allows one to evaluate the probability of observing the already recorded allelic type in new individuals to be drawn from the population, through the corresponding predictive distributions (see Section $1$ of the the Supplementary Material for more details). The package  efficiently implements the Monte Carlo approximation discussed above by exploiting the dynamic programming paradigm and optimized routines in the C++ language (via the Rcpp package, \cite{EddelbuettelFrancois2011}). The recursive structure of the problem is exploited and a suitable pruning strategy allows to save both computational time and memory requirements. Extensive documentation and a vignette are also available to increase the accessibility.

\section{Acknowledgements}

MR was partially supported by the European Union -- Next Generation EU, PRIN- PNRR 2022 (P2022H5WZ9). 
%The authors are grateful to three anonymous Referees for their helpful comments.

%\nocite*{}
\bibliographystyle{plainnat-revised}
\bibliography{reference}

\newpage

\begin{center}
\textbf{\LARGE Supplementary material }
\end{center}

\setcounter{section}{0}
\setcounter{equation}{0}
\section{Introduction}

The Hidden Markov model described in the paper can be described as
\begin{equation}\label{FVDDP_model}
Y_{t,i} |  X_t = x \overset{\text{iid}}{\sim} x, \quad X \sim \text{FV}(\alpha),\quad X_0 \sim \Pi_\alpha,
\end{equation}
where $X:=\{X_t,t\ge0\}$ is a FV process with parameter $\alpha$, reversible with respect to the law $\Pi_\alpha$ of a Dirichlet process \citep{Ferguson1973}. This FV diffusion has transition function as in $(1)$ of the main document, and its infinitesimal  generator can be written
\begin{equation}\label{eq:FVgenerator}
\mathbb{A}\phi^{m}(\mu):=\sum_{i=1}^m\langle A_if,\mu^m\rangle
+\frac{1}{2}\sum_{1\le k\ne i\le m}\langle \Phi_{ki}f-f,\mu^m\rangle,
\end{equation} 
acting on functions $\phi^{m}(\mu)=\langle f,\mu^m\rangle=\int_{\mathbb{Y}^{m}}fd \mu^{m}$, for $f\in B(\mathbb{Y}^m)$ and $\mu^{m}$ being a $m$-fold product measure. Furthermore, $A_i$ is the mutation operator
\begin{equation}\nonumber
Af(x)=\frac{1}{2}\theta\int[f(z)-f(x)]\nu_0(d z),
\end{equation}
which is the generator of a Feller process on $\mathbb{Y}$, operating on $f$ as a function of $x_j$ alone, and $\Phi_{ij}f$ is the function of $m-1$ variables obtained by setting the $i$th and the $j$th variables in $f$ equal. Cf.~\cite{EthierKurtz1993}, Section 3.

Assume data $Y_{t_k, i} \in \mathbb{Y}$ are received at times $0 = t_0 < t_1 < \dots< t_{p} = T$ and denote by $\bfY_{k} = (Y_{t_k, 1}, \dots, Y_{t_k, n_k})$ the vector of all observations collected at times $t_k$. Let also $\bfY_{0:T}=(\bfY_{0},\ldots,\bfY_{T})$ denote the entire data set. In biological applications the values contained in $\bfY_{0:T}$ are allelic types, but the theory we develop holds in general for locations in a Polish space. Denote by $(y_1^*, \dots, y_K^*)$ the vector of unique values in $\bfY_{0:T}$, with $K \in \mathbb{N}$ being their total number. Finally, we denote vectors in $\Z^K_+$ as $\bfn = (n_1, \dots, n_K)$, where $\bfm\le \bfn$ if $m_{i}\le n_{i}$ for $i=1,\ldots,K$, and let $\bfn_j \in \Z^K_+$ be the vector of multiplicities of observations in $\bfY_{j}$. Finally, set $|\bfn| := \sum_{k = 1}^Kn_k$. We collect in  Table \eqref{table:definition} a summary of the used notation.

\begin{table}
\begin{center}
\begin{tabular}{ll}
\hline
  $\mathbb{Y}$ & Set of all possible types\\
  $Y_{t, i}$ & Type of individual $i$ at time $t$\\
  $\bfY_{k}$ & Vector with all observations collected at time $t_k$\\
  $\bfY_{0:T}$   & Vector containing all the collected observations \\
  $K$ & Number of unique types in $\bfY_{[0, T]}$\\
  $y_i^*$ & $i$-th unique type in $\bfY_{[0, T]}$, with $i = 1, \dots, K$\\
\hline
\end{tabular}
%\begin{quote}
     \caption{\footnotesize  List of useful notations}
%\end{quote}
\label{table:definition}
\end{center}
\end{table}

%\begin{table}
%\caption{List of useful notations}\label{table:definition}
%\begin{tabularx}{\textwidth}{@{}XX@{}}
%\toprule
%  $\mathbb{Y}$ & Set of all possible types.\\
%  $Y_{t, i}$ & Type of individual $i$ at time $t$.\\
%  $Y_{t_r}$ & Vector with all observations collected at time $t_r$, with $r = 0, \dots, p$. \\
%  $\bfY_{[0, T]}$   & Vector containing all the collected observations. \\
%  $K$ & Number of unique types in $\bfY_{[0, T]}$.\\
%  $y_i^*$ & i-th unique type, with $i = 1, \dots, K$.\\
%\bottomrule
%\end{tabularx}
%\end{table}

The main result, which motivates the introduction of the present package, is obtained combining Proposition $3.1$ in \cite{PapaspiliopoulosRuggieroSpano2016} and Theorem $1.1$ in \cite{AscolaniLijoiRuggiero2023} and is summarized next. It states that the distribution of $X_t$, given $\bfY_{0:T}$, is always given by a finite mixture of Dirichlet processes.
\begin{theorem}\label{summary theorem}
Consider model \eqref{FVDDP_model}. Then for every $t$ there exists a set $\bfM\subset \Z_{+}^{K}$ of finite cardinality and weights $\{w_{\bfm},\bfm\in \bfM\}$ summing up to one such that 
\begin{equation}\label{summary theorem FV}
\Law(X_t|\bfY_{0:T})=\sum_{\bfm\in \bfM}w_{\bfm}\Pi_{\alpha + \sum_{k = 1}^{K}m_{k}\delta_{y^*_{k}}},
\end{equation} 
where $\Pi_\alpha$ denotes the law of a Dirichlet process with baseline measure $\alpha$.
\end{theorem}
Notice that the representation \eqref{summary theorem FV} allows also to derive the predictive distribution of new observations at time $t$, given all the available information. This follows by the nice properties of mixtures of Dirichlet processes, cf.~Corollary $3.8$ in \cite{AscolaniLijoiRuggiero2023}, whereby from \eqref{summary theorem FV} one yields the mixture of P\'olya urn schemes
\begin{equation}\nonumber%\label{}
Y_{t,1}|\bfY_{0:T},\bfm \sim 
\frac{\alpha + \sum_{k = 1}^{K}m_{k}\delta_{y^*_{k}}}{\theta + \sum_{k = 1}^{K}m_{k}}, \quad \quad \bfm|\bfY_{0:T}\sim w_{\bfm}.
\end{equation}

In the following we discuss how the weights $\{w_{\bfm},\bfm\in \bfM\}$ can be recursively computed. We consider two distinct cases:
\begin{enumerate}
\item $t \geq T$; in this case $X_t$ is conditioned only to past observations, and the left-hand side of \eqref{summary theorem FV} is the \emph{filtering distribution};
\item $t = t_j$, for $j \in {0, \dots, p-1}$; ; in this case $X_t$ is conditioned  to both data collected prior to $t$ and that become available at a later time, and the left-hand side of \eqref{summary theorem FV} is the \emph{smoothing distribution} (cf.~Section \ref{section:smoothing}).
\end{enumerate}
Notice that the case $t < T$ with $t \neq t_j$ for every $j$ can be incorporated in case 2 above without loss of generality, by adding a new collection time where no datapoints are observed.

We discuss the details about the above two cases in Sections \ref{section:predictive} and \ref{section:smoothing} respectively.
Section \ref{section:approximation} discusses how to approximate the filtering distribution with a Monte Carlo procedure implemented in the package, while Sections \ref{section:code} and \ref{section:details_figure} provide additional details on some aspects of the code and on the setting underlying Figure $1$ in the main document.

\section{Filtering distributions}\label{section:predictive}

Let $t\ge T$ in \eqref{summary theorem FV}.
The law $\Law(X_t|\bfY_{0:T})$ is obtained by recursively computing $\Law(X_{t_j}|\bfY_{0:j})$, where $\bfY_{i:j}$ denotes the set of observations collected in the time instances $(t_{i},\ldots,t_{j})$. We unpack this computation into more elementary operations, stated in the next three propositions.
\begin{proposition}\label{prop:first_update}
Consider model \eqref{FVDDP_model}. Then
\[
\Law \left(X_{t_0} |  \bfY_{0}\right) =  \Pi_{\alpha + \sum_{k=1}^K n_{0,k} \delta_{y_k^\ast}}.
\]
\end{proposition}
Proposition \ref{prop:first_update} follows immediately by the conjugacy of Dirichlet processes \citep{Ferguson1973}, and amounts to conditioning the stationary distribution to the first available data. Here $t_{0}$ is taken to be 0 without loss of generality, as by stationarity if $X_{0}\sim \Pi_{\alpha}$ then $X_{s}\sim \Pi_{\alpha}$ for any $s>0$ until the first available data point. Next we consider prediction of the signal, given past data. In relation to Proposition \ref{prop:first_update} this would mean the law of $X_{t_{0}+s}$ given $\bfY_{0}$. More generally, we condition on the previous $j$ samples to elicit the recursive nature of the computation, and show how to evaluate the law of the unobserved signal at time $t_{j}$, given information collected up to time $t_{j-1}$. 
\begin{proposition}\label{prop:propagation}
Consider model \eqref{FVDDP_model} and assume
\[
\Law \left(X_{t_{j-1}} |  \bfY_{0:j-1}\right) =  \sum_{\bfm \in \bfM}w_\bfm\Pi_{\alpha + \sum_{k=1}^K m_{k} \delta_{y_k^\ast}},
\]
for some $\bfM \subset \Z^K_+$ where $\bfM$ has finite cardinality. Then
\[
\Law \left(X_{t_{j}} |  \bfY_{0:j-1}\right) =  \sum_{\bfn \in L(\bfM)}\tilde{w}_\bfn\Pi_{\alpha + \sum_{k=1}^K m_{k} \delta_{y_k^\ast}},
\]
where 
\begin{equation}\label{lower set}
L(\bfM) = \{ \bfn \in \Z_{+}^{K} : \exists \ \bfm \in \bfM : \bfn \leq \bfm\}
\end{equation} 
and
\[
\tilde{w}_\bfn = \sum_{\bfm \in \bfM \, : \, \bfn \leq \bfm}w_\bfm p_{\bfm, \bfn}(t_{j}-t_{j-1}), 
\]
with $p_{\bfm, \bfn}(s)$ described below.
\end{proposition}
Proposition \ref{prop:propagation} amounts to Proposition $3.1$ in \cite{PapaspiliopoulosRuggieroSpano2016}. The weights $p_{\bfm, \bfn}(s)$ are  the transition probabilities of a death process that jumps from $\bfm$ to $\bfm-e_{i}$ at rate $m_{j}(\theta+|\bfm|-1)/2$, with $e_{i}$ being the canonical vector in the $i$-th direction, given by
\begin{equation}\label{eq:weights_death}
p_{\bfm, \bfn}(s)=
\begin{cases}
    e^{-\lambda_{|\bfm|}s} \quad  &\text{if} \ \bfn = \bfm   \\ C_{|\bfm|, |\bfn |}(s) \mathrm{MVH} \left(\bfn ; |\bfn|, \bfm\right) \quad &\text{if} \ \bfn < \bfm
    \end{cases}
\end{equation}
where $\lambda_{|\bfm|} = |\bfm|(\theta + |\bfm | -1)/2$, 
\begin{equation}\label{MVH}
\mathrm{MVH} \left(\bfn ; |\bfn|, \bfm\right) = \binom{|\bfm|}{|\bfn|}^{-1}\prod_{j \geq 1}\binom{m_j}{n_j}
\end{equation} 
is the probability mass function of a multivariate hypergeometric random variable and 
\[
C_{|\bfm|, |\bfn |}(s) = \left(\prod_{h=|\bfn | + 1}^{|\bfm |} \lambda_{h} \right)  (-1)^{|\bfm|-|\bfn|} \sum_{k=|\bfn |}^{|\bfm|} \frac{e^{-\lambda_{k} s}}{\prod_{|\bfn | \leq h \leq |\bfm |, h \neq k }(\lambda_{k} - \lambda_{h})}.
\]
Finally, we show how to update the distribution obtained through Proposition \ref{prop:propagation}, once further data become available at time $t_{j}$. 
\begin{proposition}\label{prop:propagation_update}
Consider model \eqref{FVDDP_model} and let
\[
\Law \left(X_{t_{j}} |  \bfY_{0:j-1}\right) =  \sum_{\bfm \in \bfM}w_\bfm\Pi_{\alpha + \sum_{k=1}^K m_{k} \delta_{y_k^\ast}},
\]
for some $\bfM \subset \Z^K_+$. Then
\[
\Law \left(X_{t_{j}} |  \bfY_{0:j}\right) =  \sum_{\bfn \in \bfM+\bfn_j}\tilde{w}_\bfn\Pi_{\alpha + \sum_{k=1}^K n_k \delta_{y_k^\ast}},
\]
where $\bfM+\bfn_j = \{ \bfm + \bfn_j : \bfm \in \bfM\}$ and
\[
\tilde{w}_{\bfm+\bfn_j} \propto w_\bfm\mathrm{PU}(\bfn_j |  \bfm),
\]
with $\mathrm{PU}(\bfn |  \bfm)$ the probability of drawing a vector $\bfn$ according to the predictive scheme
\[
\bfY_{i+1} |  \bfY_{1:i} \sim \frac{\theta P_0+\sum_{k = 1}^Km_k \delta_{y_k^*}+\sum_{i = 1}^i\delta_{\bfY_i}}{\theta+|\bfm|+i}, \quad i = 0, \dots, |\bfn|-1.
\]
\end{proposition}
Proposition \ref{prop:propagation_update} is given by Lemma $3.1$ in \cite{PapaspiliopoulosRuggieroSpano2016}, where $\text{PU}(\bfn |  \bfm)$ denotes the probability of drawing through P\'olya urn sampling a vector of multiplicities $\bfn$ from an urn with initial composition given by $\bfm$.
The above three Propositions show how the filtering distribution can be computed for every data collection time $t_{j}$. 

\begin{figure}
     \centering
     \begin{subfigure}[b]{0.24\textwidth}
         \centering
         \includegraphics[width=.8\textwidth]{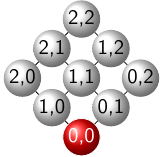}
     \end{subfigure}
     \begin{subfigure}[b]{0.24\textwidth}
         \centering
         \includegraphics[width=.8\textwidth]{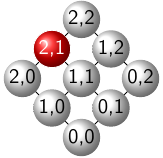}
     \end{subfigure}
     \begin{subfigure}[b]{0.24\textwidth}
         \centering
         \includegraphics[width=.8\textwidth]{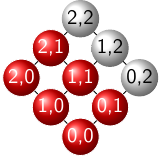}
     \end{subfigure}     
     \begin{subfigure}[b]{0.24\textwidth}
         \centering
         \includegraphics[width=.8\textwidth]{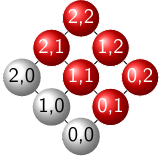}
     \end{subfigure}
          \caption{\footnotesize Graphs in $\Z_{+}^{2}$ representing the mixture components with positive weight in the current mixture of interest. From left: single mixture component when no observations are available; after observing multiplicities 2 and 1 for the two available types respectively; after propagating the signal and before further data collections; at the further data collection time after observing multiplicities 0 and 1 for the two available types.}\label{fig:graphs}
\end{figure}

Note that each of the above distributions is indexed by a finite set $\bfM \subset \Z^K_+$, and  can thus be associated to a finite graph on $\Z^K_+$, where each node corresponds to an element $\bfm\in \bfM$ and identifies a mixture component.
We clarify this intuition with a simple example in the special case of $K = 2$ possible observed types. In this case, the Dirichlet process $\Pi_{\alpha}$ reduces to a Beta distribution $\pi_{\alpha_{1},\alpha_{2}}=\text{Beta}(\alpha_{1},\alpha_{2})$. In absence of data, the prior/initial distribution $\pi_{\alpha_{1},\alpha_{2}}$ does not carry any multiplicities of types, and it is then associated to the graph origin (Figure \ref{fig:graphs}, leftmost panel). Suppose now the data at time $t_{0}$ carry multiplicities $\bfn_0 = (2,1)$, i.e.,  we have collected two data points of type 1 and one of type 2. By Proposition \eqref{prop:first_update}, the graph is now given by the single node $\bfn_0$, since the initial distribution has been updated to $\pi_{\alpha_{1}+2,\alpha_{2}+1}$ (Figure \ref{fig:graphs}, second panel). As described by Proposition \ref{prop:propagation}, when the signal is propagated from $t_{0}$ to $t_1$, the probability mass initially concentrated on $\bfn_{0}$ is spread onto lower nodes, i.e., nodes in $L(\bfn_{0})$, including the starting node (Figure \ref{fig:graphs}, third panel). How this mass is spread is governed by the death process on $\Z^K_+$ through its transition probabilities. This operation thus leads to an increase in the cardinality of the mixture components, which does not depend on the time interval considered. The mixture components are now $\pi_{\alpha_{1}+i,\alpha_{2}+j}$ with $(0,0)\le (i,j)\le (2,1)$. A further update with $\bfn_{1}=(0,1)$, as described by Proposition \ref{prop:propagation_update}, shifts the probability mass upwards by exactly $\bfn_{1}$, since each component is now $\pi_{\alpha_{1}+i+0,\alpha_{2}+j+1}$ (Figure \ref{fig:graphs}, rightmost panel).

\section{Smoothing distributions}\label{section:smoothing}

\begin{figure}
     \centering
     \begin{subfigure}[b]{0.4\textwidth}
         \centering
         \includegraphics[width=\textwidth]{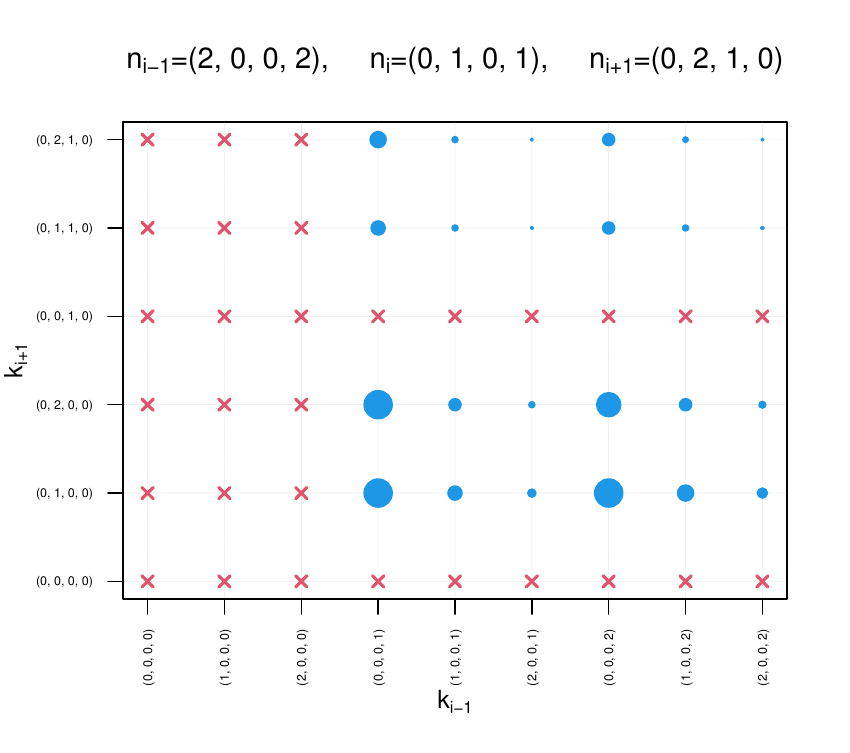}
     \end{subfigure}
     \space{1cm}
     \begin{subfigure}[b]{0.4\textwidth}
         \centering
         \includegraphics[width=\textwidth]{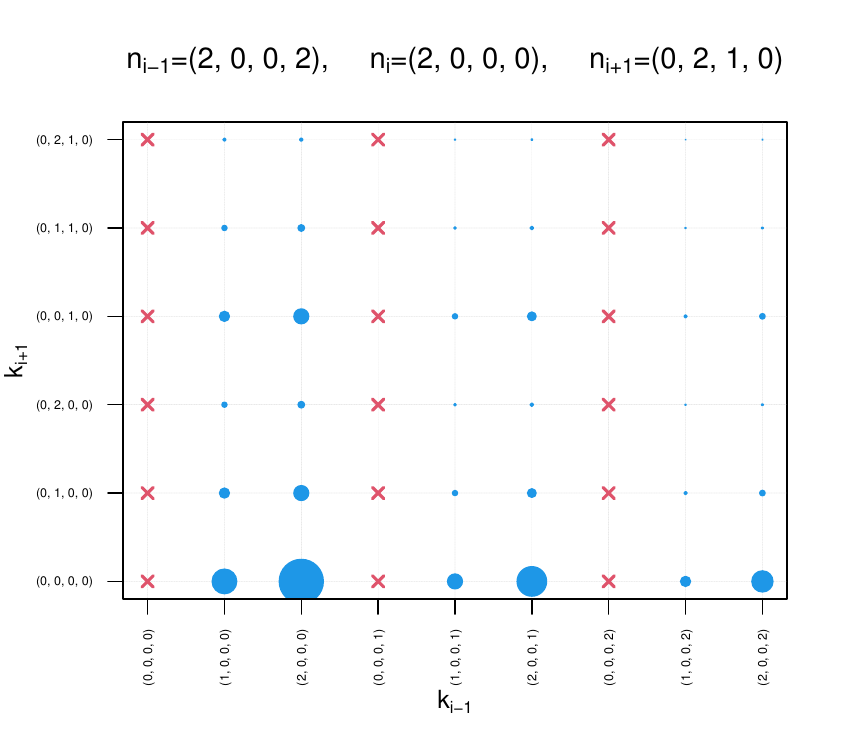}
     \end{subfigure}
          \centering
     \begin{subfigure}[b]{0.4\textwidth}
         \centering
         \includegraphics[width=\textwidth]{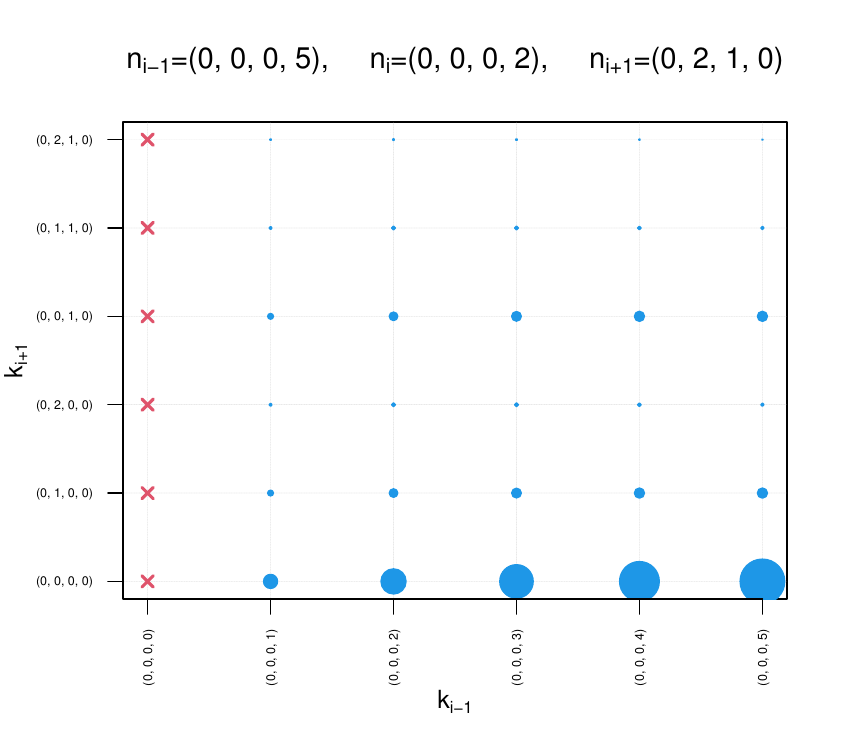}
     \end{subfigure}
     \space{1cm}
     \begin{subfigure}[b]{0.4\textwidth}
         \centering
         \includegraphics[width=\textwidth]{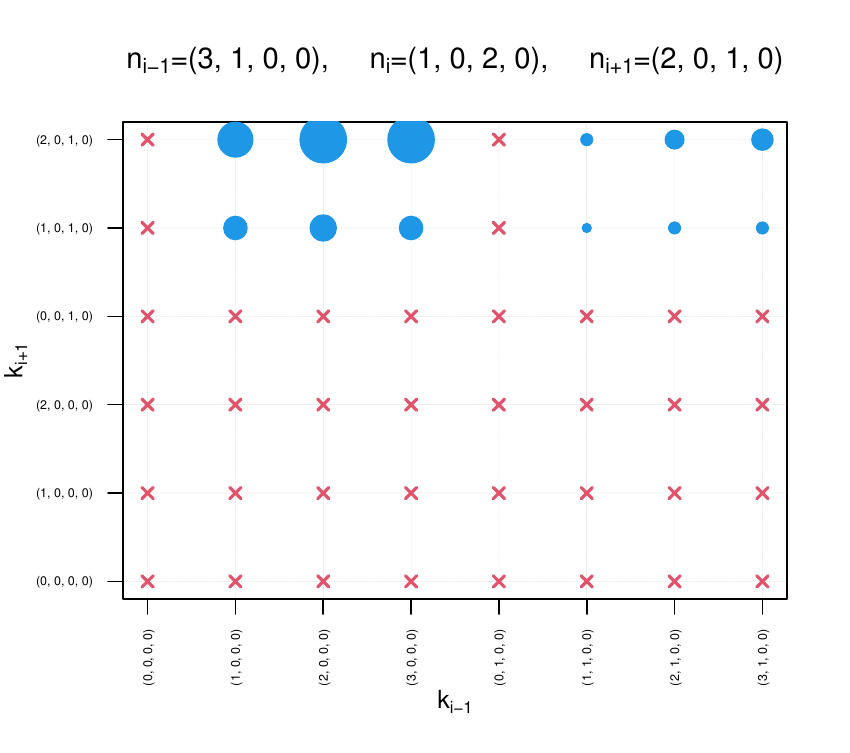}
     \end{subfigure}
               \centering
     \begin{subfigure}[b]{0.4\textwidth}
         \centering
         \includegraphics[width=\textwidth]{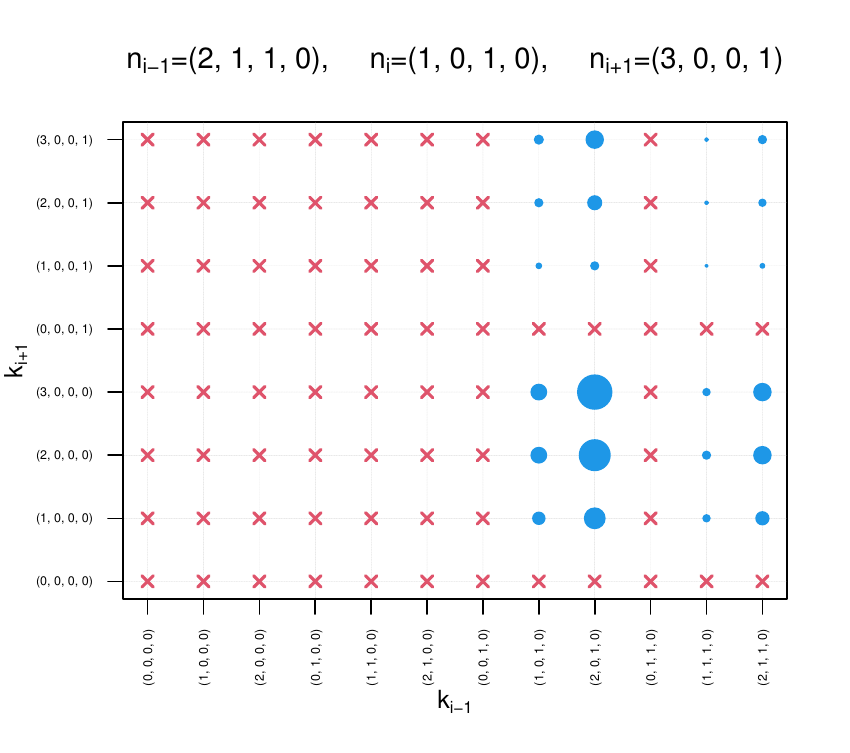}
     \end{subfigure}
     \space{1cm}
     \begin{subfigure}[b]{0.4\textwidth}
         \centering
         \includegraphics[width=\textwidth]{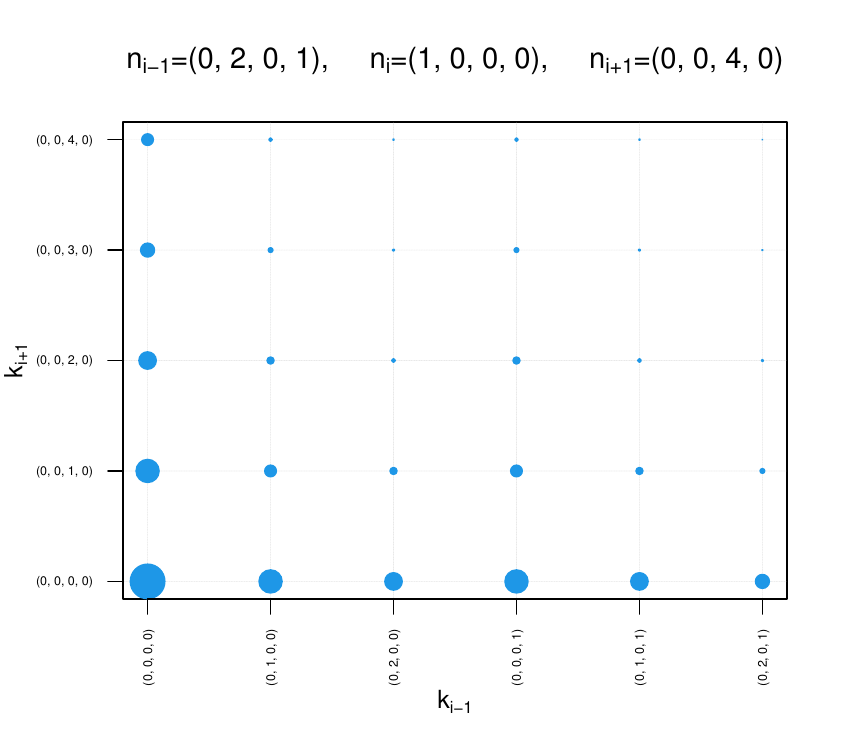}
     \end{subfigure}
     \caption{\footnotesize Weights $q(\bfkpast, \bfn_j, \bfkfuture)$ of the smoothing distribution in the case of a nonatomic $\alpha$. The blue bubbles represent the probability mass associated to each node, while red crosses denote null mass. Each plot corresponds to a different choice of $\bfm_{j-1}$ and $\bfm_{j+1}$.}\label{fig:smoothweights}
\end{figure}

Let now $t = t_j$, with $j \leq p-1$, so that the conditioning is on data collected both before and after $t_{j}$. From Section \ref{section:predictive} we know that the law of $X_{t_{j-1}}$ given $  \bfY_{0:j-1}$ can be written as
\begin{equation}\label{eq:past}
\Law \left(X_{t_{j-1}} |   \bfY_{0:j-1}\right) = \sum_{\bfm_{j-1} \in \bfM_{j-1}} u_{\bfm_{j-1}} \Pi_{\alpha + \sum_{k=1}^K m_{j-1,k}\delta_{y_k^\ast}},
\end{equation}
for some $\bfM_{j-1} \subset \Z^K_+$ of finite cardinality. By Lemma $3.2$ in \cite{AscolaniLijoiRuggiero2023} we can compute the law $\Law \left(X_{t_{j+1}} |   \bfY_{j+1:p}\right)$ in the same way, i.e., propagating backwards is the same as propagating forward. Thus, by analogy with \eqref{eq:past}, we can write
\begin{equation}\label{eq:future}
\Law \left(X_{t_{j+1}} |   \bfY_{j+1:p}\right) = \sum_{\bfm_{j+1} \in \bfM_{j+1}} v_{\bfm_{j+1}} \Pi_{\alpha + \sum_{k=1}^K m_{j+1,k}\delta_{y_k^\ast}},
\end{equation}
for some $\bfM_{j+1} \subset \Z^K_+$ of finite cardinality. The next proposition (which summarizes Proposition $3.6$ in \cite{AscolaniLijoiRuggiero2023}) shows how to compute $\Law \left(X_{t} |   \bfY_{0:T}\right)$ from \eqref{eq:past} and \eqref{eq:future}.
\begin{proposition}\label{prop:smoothing}
Consider model \eqref{FVDDP_model} and let $\Law \left(X_{t_{j-1}} |   \bfY_{0:j-1}\right)$ and $\Law \left(X_{t_{j+1}} |   \bfY_{j+1:p}\right)$ be as in \eqref{eq:past} and \eqref{eq:future}. Then
\begin{align*}
\Law\left(X_{t} |  \bfY_{0:T}\right) =  & \sum_{\substack{\bfmpast \in M_{j-1}}}\sum_{\substack{\bfmfuture \in M_{j+1}}}  u_{\bfmpast} v_{\bfmfuture} f(\bfmpast,\bfmfuture),%\\ 
%& \sum_{\substack{(\bfkpast, \bfkfuture) \, : \, \\ \bfkpast \leq \bfmpast, \bfkfuture \leq \bfmfuture}} \tilde{p}_{\bfkpast, \bfkfuture}^{\bfmpast, \bfmfuture}q(\bfkpast, \bfn_j, \bfkfuture)\Pi_{\alpha  + \sum_{k=1}^K (\bfk_{j-1, k} + \bfn_{j,k} + \bfk_{j+1, k} )\delta_{y_k^\ast}},
\end{align*}
where 
\begin{equation}\nonumber%\label{}
\begin{aligned}
f(\bfmpast,\bfmfuture)=&\, \sum_{\substack{(\bfkpast, \bfkfuture) \, : \, \\ \bfkpast \leq \bfmpast, \bfkfuture \leq \bfmfuture}} \tilde{p}_{\bfkpast, \bfkfuture}^{\bfmpast, \bfmfuture}q(\bfkpast, \bfn_j, \bfkfuture)\Pi_{\alpha  + \sum_{k=1}^K (\bfk_{j-1, k} + \bfn_{j,k} + \bfk_{j+1, k} )\delta_{y_k^\ast}}\\
\tilde{p}_{\bfkpast, \bfkfuture}^{\bfmpast, \bfmfuture} =&\, p_{\bfmpast, \bfkpast}(t_j - t_{j-1})p_{\bfmfuture, \bfkfuture}(t_{j+1} - t_j)
\end{aligned}
\end{equation} 
with $p_{\bfm, \bfn}(s)$ as in \eqref{eq:weights_death} and $q(\bfkpast, \bfn_j, \bfkfuture)$ as given below.
\end{proposition}
By readjusting the weights, it is then easy to obtain the representation (5) in the main document. The explicit formula for the weights $q(\bfkpast, \bfn_j, \bfkfuture)$ depends on the baseline measure $\alpha$, as given by Proposition $3.6$ in \cite{AscolaniLijoiRuggiero2023}. In particular:
\begin{itemize}

\item if $\alpha$ is \emph{atomic} on $(y_1^*, \dots, y_K^*)$, these can be expressed as
\[
 q(\bfkpast, \bfn_j, \bfkfuture)\ \propto  \frac{m(\bfkpast + \bfn_j +\bfkfuture)}{m(\bfkpast)m(\bfn_j)m(\bfkfuture)}
 \]
with
\[
m(\bfn) = \frac{B(\bfalpha+\bfn)}{B(\bfalpha)},\quad 
B(\bfalpha):=\frac{\prod_{k=1}^{K}\Gamma(\alpha_{k})}{\Gamma(|\bfalpha|)}, \quad \bfalpha = (\alpha(y_1^*), \dots, \alpha(y_K^*)).
\]

    \item if $\alpha$ is \emph{nonatomic}, define the sets
$$ \mathcal{D}_{j-1}^{\bfmpast,\bfmfuture} := \{ k \in \{ 1, \dots, K\} : m_{j-1, k} > 0 \ \text{and either} \ n_{j,k}>0 \ \text{or} \  m_{j+1,k}>0 \},$$
$$\mathcal{D}_{j+1}^{\bfmpast,\bfmfuture} := \{ k \in \{ 1, \dots, K\} : m_{j+1, k} > 0 \ \text{and either} \ n_{j,k}>0 \ \text{or} \  m_{j-1,k}>0 \}$$
and $$
\mathcal{S}^{\bfnpast,\bfnfuture} := \mathcal{D}_{j-1}^{\bfnpast,\bfnfuture} \cup \mathcal{D}_{j+1}^{\bfnpast,\bfnfuture}$$
to express the indices of shared values among different times. Then
\begin{align*} \mathcal{D}^{\bfmpast,\bfmfuture} = \{ (\bfkpast, \bfkfuture) : &\bfkpast \leq \bfmpast \ \text{and} \ k_{j-1, k} > 0 \ \forall \ k \in \mathcal{D}^{\bfmpast,\bfmfuture}_{j-1},\\
    & \bfkfuture \leq \bfmfuture \ \text{and}  \ k_{j+1, k} > 0 \ \forall \ k \in \mathcal{D}^{\bfmpast,\bfmfuture}_{j+1} \} \end{align*}
    and the weights are such that:
    \begin{itemize}
    \item if $\mathcal{S}^{\bfmpast,\bfmfuture} = \emptyset$:
    $$q(\bfkpast, \bfn_j, \bfkfuture)\ \propto  \frac{\theta^{(|\bfkpast|)} \theta^{(|\bfkfuture|)}}{(\theta + |\bfn_j|)^{(|\bfkpast|+|\bfkfuture|)}}$$
    \item if $\mathcal{S}^{\bfmpast,\bfmfuture} \neq \emptyset$:
\begin{align*}
q(\bfkpast, \bfn_j, \bfkfuture) \ \propto &\frac{\theta^{(|\bfkpast|)} \theta^{(|\bfkfuture|)}}{(\theta + |\bfm_j|)^{(|\bfkpast|+|\bfkfuture|)}} \cdot \\   & \cdot \prod_{k \in \mathcal{S}}\frac{(k_{j-1, k} + n_{j,k} + k_{j+1,k}-1)!}{(k_{j-1,k}-1)! (n_{j,k}-1)! (k_{j-1,jk}-1)!} \end{align*}
if $(\bfkpast, \bfkfuture) \in \mathcal{D}^{\bfmpast,\bfmfuture}$, and $0$ otherwise, under the convention that $(-1)!=1$.
\end{itemize}
\end{itemize}
Notice that, when $\alpha$ is nonatomic and a type $y_k^\ast$ is observed at different times, the smoothing distribution imposes that the probability mass is concentrated on components of the mixture in which the type $y_k^\ast$ is available (i.e., $m_k > 0$). This implies that the smoothing operation automatically implies a pruning of the graph, eliminating nodes which do not carry shared information available in the data. This implication is illustrated  in Figure \ref{fig:smoothweights}. The Figure displays  the weights $q(\bfkpast, \bfn_j, \bfkfuture)$ (as bubbles) associated to each pair $(\bfkpast, \bfkfuture)$ (sorted along the axes) for different combinations of $\bfmpast$, $\bfm_j$ and $\bfmfuture$. In particular the figure shows how the probability masses for certain nodes are exactly zero after the update (crosses), typically near one or more sides of the lattice or, as discussed above, when the information is not shared among different collection times. The probability mass also appears to vary strongly among the nodes, which suggests many nodes may carry very small if not negligible probability.
\\

\section{Approximating filtering and smoothing distributions}\label{section:approximation}

Given the representation discussed at the end of Section \ref{section:predictive}, each distribution of interest is characterized by a graph on $\Z^K_+$ and the corresponding weights. By Proposition \ref{prop:propagation} each propagation step increases the number of elements in the graph and it can be shown that the latter grows polynomially with the number of collected datapoints (see, e.g., Section $3.2$ of \cite{AscolaniLijoiRuggiero2023}). This phenomenon makes exact computation of the weights impractical even with datasets of moderate size. However, as discussed in the main document, it is possible to exploit  the sparsity properties of the involved distributions to devise an efficient Monte Carlo approximation. In the Section ``Methods'' of the main document it is shown how to approximate the weights of the smoothing distribution starting from the (forward and backward) filtering distributions in \eqref{eq:past} and \eqref{eq:future}. Here we complete that discussion 
by showing how to approximate the filtering distributions.

Notice that updating the finite mixture with current observations (see Proposition \ref{prop:propagation_update}) leaves the number of nodes in the graph unchanged. See third and fourth panels in Figure \ref{fig:graphs}. Thus, it suffices to deal with the propagation step (see Proposition \ref{prop:propagation}). Assume that
\[
\Law \left(X_{t_{j-1}} |  \bfY_{0:j-1}\right) \approx  \sum_{\bfm \in \bfM}w_\bfm\Pi_{\alpha + \sum_{k=1}^K m_{k} \delta_{y_k^\ast}},
\]
for some $\bfM \subset \Z^K_+$, where here $\bfM$ and $\{w_\bfm\}_{\bfm \in \bfM}$ are the graph and weights which approximate the distribution of the signal before propagating from $t_j$ to $t$. Then, by Proposition \ref{prop:propagation}, it is required to approximate $L(\bfM)$ as in \eqref{lower set} and the corresponding weights. Following the same argument of the main document, the algorithm repeats the following procedure:
\begin{itemize}
\item draw $\bfm \in \bfM$ with probability $w_{\bfm}$;
\item simulate a death process on $\Z^K_+$ starting from $\bfm$ with transition probabilities as in \eqref{eq:weights_death};
\item return the arrival node $\bfk$.
\end{itemize}
The death process simulation can proceed as follows. When in $\bfm$, and until the sum of the non-homogeneous Exponential waiting times exceeds the desired interval:
\begin{itemize}
\item draw $S\sim \text{Exp}(\lambda_{|\bfm|})$;
\item jump to $\bfn$ drawn from \eqref{MVH}, with $|\bfn|=|\bfm|-1$.
\end{itemize}
The above simulation should be performed for a desired number of particles, for every given $\bfn$, the new weight $\tilde{w}_\bfn$ is approximated by the proportion of iterations where the node $\bfn$ is the arrival point. Moreover, nodes with an empirical weights smaller than some threshold $\varepsilon$ (typically between $10^{-9}$ and the machine epsilon of the computer in use) are eliminated with a pruning procedure. 

Thus, the final output of the algorithm is given by nodes and weights $\tilde{\bfM}$ and $\{\tilde{w}_\bfn\}_{\bfn \in \tilde{\bfM}}$ such that
\[
\Law \left(X_{t_{j}} |  \bfY_{0:j-1}\right) \approx  \sum_{\bfn \in \tilde{\bfM}}\tilde{w}_\bfn\Pi_{\alpha + \sum_{k=1}^K n_{k} \delta_{y_k^\ast}}.
\]
This procedure typically leads to manageable graphs, i.e., where $\tilde{\bfM}$ has a moderate cardinality, with a large saving of computational resources. Cf.~Figure $1$ in the main document. 
%Indeed, as discussed there, a large number of weights in the exact mixture has low cumulative probability (cf. Figure $4$ in \cite{AscolaniLijoiRuggiero2021}).

\section{Details of the code}\label{section:code}
The process is implemented as an R "S3" class, i.e., as a list containing specific elements. The fixed ones, to be specified during the initialization, are necessary to describe the parameter $\alpha$ of the model; this is done expressing it as $\alpha =\theta P_0$ where $\theta$ is a positive real number and $P_0$ a probability distribution on the desired state space $\mathbb{Y}$. When the signal is updated and propagated, the relevant information is expressed via $y^\ast$ (a vector), $M$ (a matrix, where each $\bfm$ is a row) and $w$ (a vector). \\ 
Since the majority of the operations discussed above may require a large computational cost, most of them have been implemented in C++ via the Rcpp package \citep{EddelbuettelFrancois2011}. This helped in the use of dynamic programming to alleviate the execution times and the use of memory: for example, as $C_{|\bfm|,|\bfn|}(t)$ only depends on the cardinality of the vectors, the values of its calculation can be stored and retrieved in an appropriate array. Since the function that computes $ C_{|\bfm|,|\bfn|}(t)$ is used several times while iterating on the elements of the matrix $M$, it is important that the array where the results are stored is passed by reference every time that the function is called.\\
To calculate the smoothing distribution, a binning strategy is pursued: each ancestor pair $(\bfnpast, \bfnfuture)$, transformed into a single string, is used as a key for a map, which represents a bin. In turn, such bin consists of a map: in this case, the key is the descendant pair $(\bfkpast, \bfkfuture)$ and the associated values are two: the first is the weight $q(\bfkpast, \bfn_j, \bfkfuture)$, computed analytically, while the second counts how many times the pair $(\bfkpast, \bfkfuture)$ has been drawn from the ancestors $(\bfnpast, \bfnfuture)$. When all samples have been obtained, within each bin, the weights $q(\bfkpast, \bfn_j, \bfkfuture)$ referred to each pair of descendants are multiplied by the draw counts, and the resulting weights are then normalized to sum up to one. Finally, the weights are merged: the weight of the node $\bfkpast + \bfn_j + \bfkfuture$ is computed as the sum of the weights of the pair $(\bfkpast, \bfkfuture)$ in each bin, multiplied by the weight of the bin $u_{\bfnpast} \cdot v_{\bfnfuture}$.\\
This structure implies that each bin is implemented as a map whose items are themselves maps, and consequently the variable that counts how many times each node is drawn and the weights $q(\bfkpast, \bfn_j, \bfkfuture)$ can be accessed with constant cost. This also means that once the samples have been drawn, reordering them in the matrix $M$ and generating the vector of weights $w$ has a linear cost with respect to the amount of nodes generated by the approximating algorithm. This is also true regarding propagation.\\
Simpler operations, such as updating and sampling, have been implemented in R by taking advantage of the vectorization properties it provides. \\
The code is available at
\href{https://github.com/StefanoDamato/FVDDPpkg}{https://github.com/StefanoDamato/FVDDPpkg}.

\section{Details of Figure $1$ in Main Document}\label{section:details_figure}

\begin{figure}
  \centering
  \begin{subfigure}[b]{0.495\textwidth}
    \includegraphics[width=\textwidth]{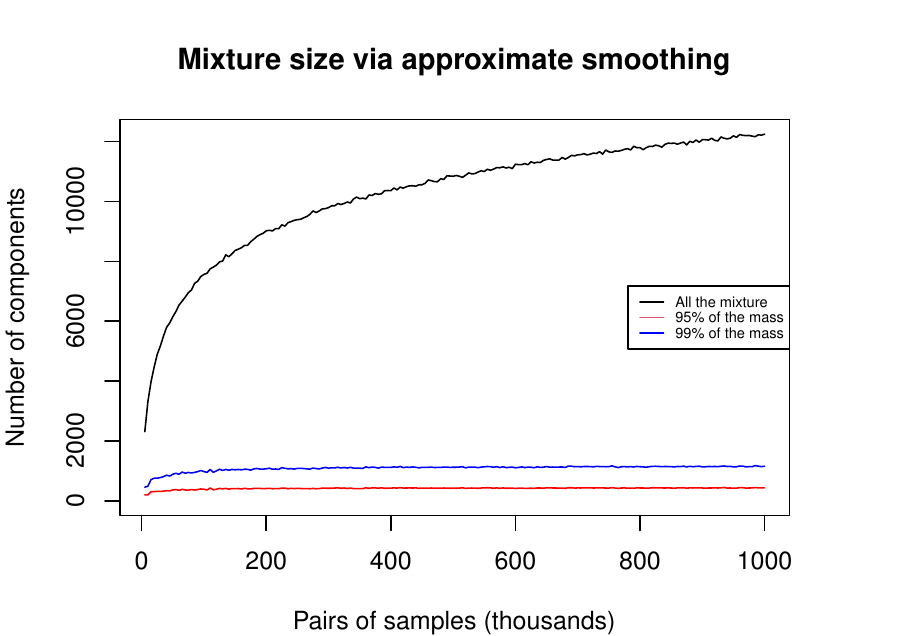}
  \end{subfigure}
  \hfill
  \begin{subfigure}[b]{0.495\textwidth}
    \includegraphics[width=\textwidth]{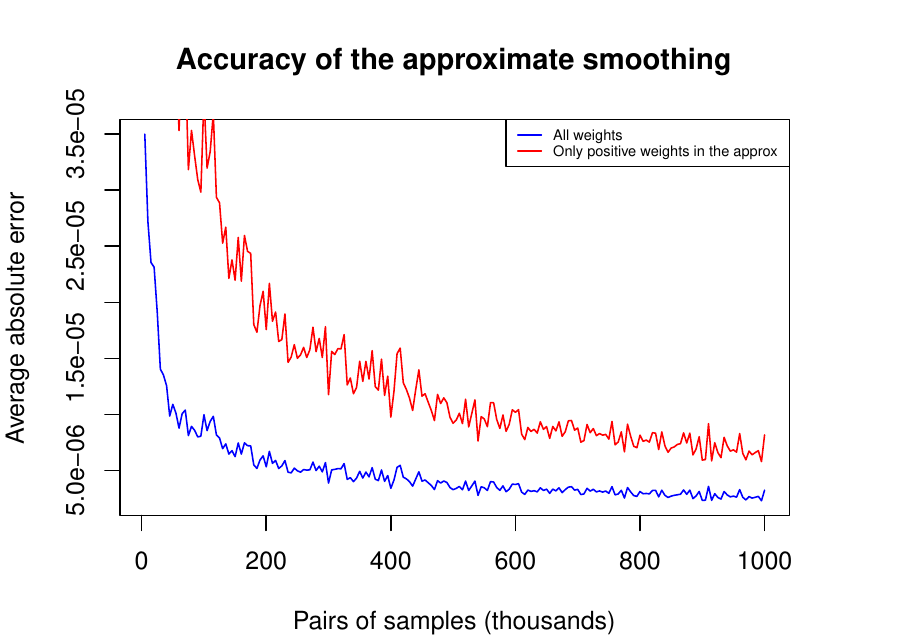}
  \end{subfigure}
  \begin{subfigure}{0.495\textwidth}
    \includegraphics[width=\textwidth]{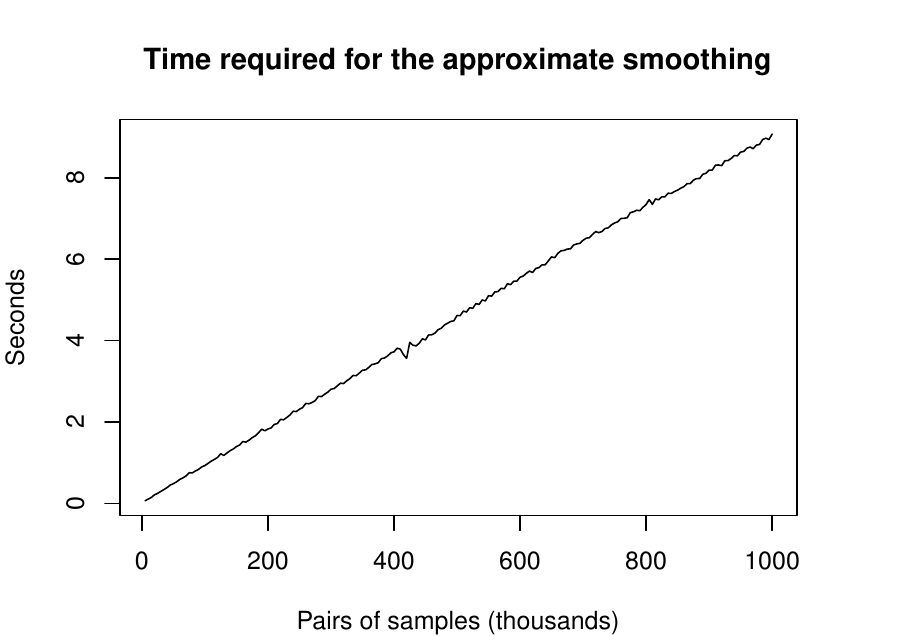}
  \end{subfigure}
       \caption{\footnotesize In the first panel, the number of components in the approximate process for different choices of the amount of Monte Carlo samples from the past and the future. In the second panel, the accuracy of the approximate weights for the same processes, where the red curve considers only nodes which were sampled at least once in the Monte Carlo approximation. In the third panel, the time required for the approximate algorithm with respect to the amount of samples. Here the data are generated according to model \eqref{model_simulation}.}
  \label{fig:approx_time_accuracy}
\end{figure}

The data on which the Figure 1 in Main Document is based were generated according to the following hierarchical model:
\begin{equation}\label{model_simulation}
 \begin{aligned}
    &\bfY_i \iid \frac12 \mathrm{Pois}(\mu_i^{-1}) + \frac12 \mathrm{Pois}(5 + \nu_i^{-1}), \\
    &\mu_{i} = \mu_{i-1} + \varepsilon_i, \quad \varepsilon_i \iid \mathrm{Exp}(1), \\
    &\nu_{i} = \nu_{i-1} + \eta_i, \quad \eta_i \iid \mathrm{Exp}(1),
\end{aligned}
\end{equation}
where $\mu_0^{-1} = \nu_0^{-1} = 5$ and $\mu_i , \nu_i$ are independent. Data were collected at times $t_0=0$, $t_1 = 0.5$ and $t_2 = 1$,  generating vectors of $10$ observations for each time, denoted $\bfY_0,\bfY_1,\bfY_2$. This is the same setting used in Section $4$ of \cite{AscolaniLijoiRuggiero2021}, where the FV specification is shown to outperform other models for time dependent data. \\
Thus, the FV model is applied to the generated dataset, by choosing $\alpha \sim \mathrm{NegBin}(2, 0.5)$ and the smoothing distribution is obtained by means of the appropriate algorithms outlined above, with propagation times $t_2 - t_1 = t_1 - t_0 = 0.5$. In particular the approximate algorithm is run drawing $10^6$ Monte Carlo samples (this operation required less than 10 seconds on our machine). \\
Figure \ref{fig:approx_time_accuracy} shows that as the number of Monte Carlo samples used to approximate the mixture grows, the amount of mixture components it contains tends to settle at a much lower value than that of the process calculated with the exact algorithm (that is 30000). The same is true for the amount of components containing $95\%$ and $99\%$ of the total probablity mass. 
%In particular, the approximated mixture is significantly smaller than the exact one, since the latter contains precisely 31104 components.
%The average error in approximating the weights stabilized after a substantial number of samples are drawn. In general, the average error of the approximate weights is larger when the exact distribution to be reproduced has an higher amount of components. 
%The Figure also shows the evolution of the average absolute error with respect to the unobserved signal when using the exact and the approximating mixture. \\
The second panel of the Figure shows how as the number of samples increases, the mixture generated by the approximate algorithm (in red) gets more accurate, as the difference of the weights from the exact one (in blue) decreases. In particular, the red curves considers only the nodes which have been sampled at least once in the Monte Carlo approximation.\\
The memory usage of the approximate algorithm, being proportional to the number of nodes in the resulting mixture, stabilizes at values that are significantly lower than those required by the exact algorithm. Figure \ref{fig:approx_time_accuracy} (first panel) depicts the effect of pruning, which significantly reduces the number of components to retain.

%\bibliographystyle{plainnat-revised}
%\bibliography{reference}

\end{document}